# Enhancing Kubernetes Automated Scheduling with Deep Learning and Reinforcement Techniques for Large-Scale Cloud Computing Optimization

Zheng Xu[1]* Yulu Gong[1], Yanlin Zhou[2], Qiaozhi Bao[3], Wenpin Qian[4]

1* Computer Engineering, Stevens Institute of Technology, hoboken, NJ, USA
1 Computer&InformationTechnology, NorthernArizonaUniversity, Flagstaff, AZ, USA
2 Computer Science, Johns Hopkins University MD, USA
3 Statistics, North Carolina State University, NC State University, Raleigh, North Carolina 27695
4 Information Science, Trine University Phoenix, Arizona, USA
*Corresponding author:[Zheng Xu, E-mail:zhengxu1994620@gmail.com]

## ABSTRACT

With the continuous expansion of the scale of cloud computing applications, artificial intelligence technologies such as Deep Learning and Reinforcement Learning have gradually become the key tools to solve the automated task scheduling of large-scale cloud computing systems. Aiming at the complexity and real-time requirement of task scheduling in large-scale cloud computing system, this paper proposes an automatic task scheduling scheme based on deep learning and reinforcement learning. Firstly, the deep learning technology is used to monitor and predict the parameters in the cloud computing system in real time to obtain the system status information. Then, combined with reinforcement learning algorithm, the task scheduling strategy is dynamically adjusted according to the real-time system state and task characteristics to achieve the optimal utilization of system resources and the maximum of task execution efficiency. This paper verifies the effectiveness and performance advantages of the proposed scheme in experiments, and proves the potential and application prospect of deep learning and reinforcement learning in automatic task scheduling in large-scale cloud computing systems.
**Key Words**：Cloud computing system; Reinforcement learning; Deep learning; Perceptual scheduling; large-scale

## 1. INSTRUCTION

With the development of cloud computing, big data and artificial intelligence (AI) technology and the industry's comprehensive application demand for these three technologies, there is a technological development trend of big data, cloud computing and artificial intelligence integration at home and abroad. The Big Data Expert Committee of the China Computer Society pointed out in the 2019 Big Data development trend survey report that artificial intelligence, big data and cloud computing will be highly integrated into an integrated system. The birth of the Hadoop project in 2006 marked the beginning of the era of big data technology, and the commercialization of Amazon web services (AWS) marked the formal step of cloud computing to change the information age. Since then, big data and cloud computing have become very hot technologies in recent decades, and academia and industry have vigorously invested in related technology research and development and landing applications, and have created several classic works of collaboration between academia and industry. As a powerful computing and storage mode, cloud computing provides strong support for the processing of big data, but it also brings a series of new challenges. In the cloud computing environment, how to efficiently store and process big data has become a topic of great concern. Optimizing big data storage and processing not only impacts the economic benefits of enterprises but also influences data-driven decision-making and innovation capabilities. In this context, this paper aims to deeply analyze the optimization strategy of big data storage and processing in cloud computing environment, with a view to providing practical guidance and enlightenment for researchers, engineers and decision makers

Hair. For example, the new computing engine Spark technology launched by the University of California, Berkeley in 2012 was quickly accepted by the industry and became a new generation of standard computing engine. In the field of cloud computing, more enterprise-level applications have promoted the development of technologies, such as container technology and orchestration system, which began to rise in 2014, and finally promoted the rapid development of a new generation of cloud native platforms, and Docker and Kubernetes technologies have become the standards of a new generation of cloud native platforms. The development of basic software is not a simple function accumulation, it must go through detailed architecture design and function verification.

Both big data and AI computing are typical examples of distributed computing models, relying on directed acyclic graph (DAG) or massive parallel programming (MPP) iterations. It means that computing tasks are generated at runtime, so it is difficult to schedule in advance, and the distributed characteristics require higher flexibility and adaptability of scheduling systems. At present, the industry is trying to deploy big data platforms and AI technologies on cloud native to achieve greater resilience and scalability. However, there have been few breakthroughs in this field on a global scale. This article will cover core innovation: how core scheduling systems in cloud platforms manage and schedule big data and AI applications on cloud native platforms.

## 2. TASK SCHEDULING PROBLEM IN CLOUD COMPUTING SYSTEM

### 2.1 Cloud computing

Cloud computing refers to the use of the cloud as its production deployment mode during application development, so as to make full use of the core advantages of the cloud such as elasticity, scalability, and self-healing. Different from the traditional bloated single application development mode, cloud-native application has become the current mainstream application development mode because of its effective collaborative development, testing and operation and maintenance, which greatly improves the efficiency and quality of software development and supports rapid product launch and iteration.

The platform that can effectively support cloud native applications is usually called cloud native platform, which is mainly characterized by containerized packaging, automated management and microservices-oriented system. Docker directly uses LinuxCgroup and other technologies to provide logical virtualization. Docker has become the preferred application container technology because of its features such as low system overhead, fast startup, good isolation, and convenient application encapsulation.

## 2.2 Cloud computing system architecture

The cloud computing system as a whole is divided into two parts: front-end and back-end. These two parts are connected to each other through a network called the Internet. The front-end is the computer user or client, and the back-end system is the so-called "cloud" server.

The front-end is for customers to access their own applications on the cloud computing system through their own computer networks, and the cloud computing system provides corresponding user interfaces according to the needs of customers. For example, web-based email systems use existing Web browsers such as Internet Explorer or Foxmail to access mail systems on "cloud" servers. Other unique applications can receive professional-grade services from the "cloud" server through a specific client application.

The back-end is a variety of computer server systems and data cloud storage systems based on the Internet to provide customers with a variety of "cloud" services. In theory, a cloud computing system can include almost all computer programs, from data processing to video games, and each application has its own dedicated server, which is maintained and managed by a professional team.

All "cloud" servers are subject to the central server management, the central server according to the customer and the network smooth situation to determine which customer to use the "cloud" server services, to ensure the smooth use of customers, it follows a set of rules that is the protocol, the use of a special software called middleware. Middleware allows networked computers to communicate with each other according to the license, through professional software, running multiple virtual servers on a physical server, each virtual server running its own independent operating system, is a set of open infrastructure virtualization platform in the cloud computing environment, server virtualization reduces the need for more physical machines.

If a cloud computing company has a lot of customers, it may need a lot of storage space, the customer stores a lot of data on the "cloud" server, the cloud computing system requires storage equipment is at least twice the storage capacity, need to store the customer's data at least two or more copies. In this way, if one storage server goes down, other servers can continue to provide services without any impact on customers.

## 2.3 Machine learning and artificial intelligence

Distributed computing is rapidly iterated and developed under the requirements of complex industrial applications, starting from the MapReduce5 calculation model for offline processing. The online computing models represented by Spark (Spark, Tez Щ u, Druidn2, etc.) and real-time computing models (Flinku3l, Spark Streamingu4, etc.) were gradually developed. The new distributed computing model opens up new application areas, but also poses great challenges to the management, scheduling and operation and maintenance of large-scale systems. As a result, big data provides rich data resources, while cloud computing platforms provide high-performance computing resources and storage capabilities that enable machine learning algorithms to analyze data more precisely, train models, and achieve intelligent decisions. For example, natural language processing applications can leverage big data for language model training in a cloud computing environment to improve the accuracy of text processing. Image recognition applications can also achieve faster image classification and recognition through the parallel computing power of cloud computing. This integration offers endless possibilities for innovative intelligent applications that help improve the user experience and increase productivity. With the accumulation of big data, data security and privacy issues have become particularly important.

In the environment of cloud computing and big data convergence, organizations need to adopt strict data encryption, access control and monitoring measures to ensure the security of large-scale data. Data breaches can lead to serious financial and reputational damage, so protecting the security of data has become an urgent task.

**2.4 Kubernetes system**

The kube-scheduler in the Kubernetes cluster monitors newly created Pods without assigned nodes and selects the most suitable node for deployment. So how do these actions or these principles work, let's take a look at them. For newly created Pods or other unscheduled Pods, kube-scheduler selects an optimal node for them to run on. However, each container in a Pod has different requirements for resources, and each Pod has different requirements. Therefore, existing nodes need to be filtered according to specific scheduling requirements.

**2.5 Kubernetes scheduling procedure**

In a Kubernetes cluster, nodes that meet Pod scheduling requirements are called feasible nodes (FN). If there is no suitable node, the pod will remain unscheduled until the scheduler can place it. That is, when we create a Pod, if it is in a Pending state for a long time, it is time to see if your cluster scheduler has no suitable nodes due to some problem

After the scheduler finds FN for the Pod, it then runs a set of functions to score FN and finds the node with the highest score in FN to run the Pod.

Factors that need to be considered in scheduling policy decision-making include individual and collective resource requirements, hardware/software/policy constraints, affinity and anti-affinity specifications, data locality, inter-workload interference, and so on.

Part One - Kubernetes scheduling process. As shown in the following figure, a simple Kubernetes cluster architecture is drawn, which includes a kube-ApiServer, a set of webhooks controllers, and a default kube-Scheduler. There are also two physical machine nodes, Node1 and Node2, on which two Kubelets are deployed.

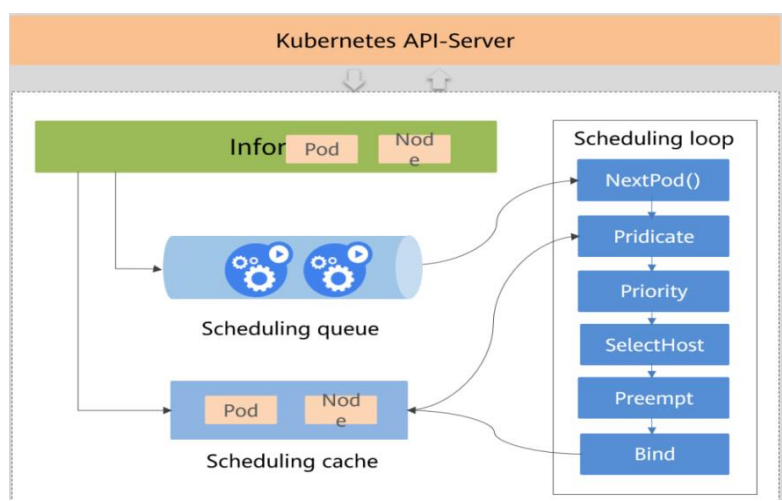

**Figure 1:**Kubernetes scheduling process architecture

The dashed lines in Figure 1 represent the main components of Kubernetes, including Informer, the scheduling queue, the scheduler's cache, and the scheduling master loop. Informer uses the list/watch mechanism to obtain resource information changes and update queue and cache. . · NextPod() obtains the Pod of the head of the queue from the queue to be scheduled;

| Algorithm Name | Feature |
|---|---|
| GeneralPredicates | It includes three basic checks: nodes, ports, and rules. For example, the maximum number of Pod resource objects on a node and whether resources such as CPU, MEM, and GPU meet requirements |
| NoDiskConflict | Check whether the Node meets the hard disk requirements of the Pod. For example, check whether the volume used by the Pod conflicts with the volume used by other Pods on the node |
| CheckVolumeBinding | Check whether the node meets the PVC mounting requirements for Pod resource objects |
| NoVolumeZoneConflict | If a single cluster is deployed across AZS, check whether the PVC to which the Pod resource object is attached is mounted across regions |
| CheckNodeMemoryPressure | Checks whether Pod resource objects can be scheduled to MemoryPresure nodes |
| CheckNodePIDPressure | Check whether the Pod resource object can be scheduled to the node with PIDPressure |
| PodToleratesNodeTaints | Check that the Pod can tolerate all taints on node |
| CheckNodeMemoryPressure | When Pod QoS is besteffort, check the remaining memory of nodes and exclude nodes with excessive memory pressure |
| MatchInterPodAffinity | Check whether the node meets the affinity and anti-affinity requirements of the pod |

. Get the Node list from the cache;

Predicate algorithm is executed for Pod and NodeList to filter out inappropriate nodes. Implement Priority algorithm for Pod and NodeList to score nodes;

According to the score, the node with the highest score is calculated:

**TABLE 1.** Name of the cloud computing scheduling algorithm

When a Pod with a higher priority does not find a suitable node, the scheduler tries to preempt a Pod with a lower priority for it. When the scheduler selects a suitable node for the Pod, Bind the Pod to the node via bind.

**Features of Kubernetes default scheduler:**

(1) Queue-based scheduler

(2) Dispatch only one at a time

(3) Pod scheduling time is globally optimal

**2.6 Kubernetes aware scheduling and policy**

Kubernetes has two kinds of algorithms for aware scheduling :Predicate and Priority. Predicate is to filter all nodes and filter out unqualified nodes, while Priority is to grade the nodes filtered by the Predicate and select the best nodes. Predicate policies filter nodes that meet the conditions. Different Pods on Node will have resource conflicts. Predicate aims to avoid resource conflicts, node overload, port conflicts, etc.

**2.7 Challenges for cloud-native batch computing**

(1) Lack of job management · PoD-level scheduling, unable to perceive the upper application. Lack

of operational concept, lack of sound life cycle management. Lack of task dependency and job dependency support

2) Scheduling policy limitations. Gang-Scheduling and Fairshaing scheduling are not supported. Resource reservation for multiple scenarios is not supported, backfill. CPU/l0 topology based scheduling is not supported

Insufficient support for domain computing frameworks

(3) · 1:1 operator deployment operation and maintenance complex Different frameworks have different requirements for job management and parallel computing. Computationally intensive, large resource fluctuations require advanced scheduling capabilities.

4) Insufficient support for resource planning reuse and heterogeneous computing, lack of queue concept

Dynamic planning of cluster resources and resource overcommitment are not supported. Heterogeneous resources are not supported.

## 3. PROPOSED METHODOLOGY FOR TASK SCHEDULING

With the development of cutting-edge technologies such as AI large models and machine learning, the demand for computing resources for AI applications is increasing day by day. The Kubernetes-based AI computing power platform can effectively integrate the underlying hardware resources, achieve fine management and optimal allocation of computing power resources, and promote the promotion and popularization of AI applications. As the world's top three contributors to the Kubernetes project, DaoCloud actively promotes the optimization of scheduling technology, improves the efficiency of resource allocation, and initiates the open source KWOK (Lightweight large-scale Cluster Simulator). Deeply involved in the research and development of Kueue (Kubernetes native job queue management system), and launched a new generation of AI computing power platform to help enterprises improve the utilization rate of computing power and promote the development and application of AI technology.

**3.1 Overview of the Proposed Scheme**

(1) Gang scheduling

Typically, when automated scheduling is implemented, a job contains multiple instances that need to be started and ended together. For the scheduler, a job cannot be scheduled until there are enough resources to schedule all instances of the job. Gang scheduling solves the deadlock problem caused by multiple tasks waiting for resources at the same time. Therefore, the community current solution abstracts the concept of a PodGroup. Both the batch scheduler Volcano and the native scheduler coscheduling plug-ins are based on this concept to implement gang scheduling.

(2) Task scheduling

For multi-task, the concept of task queue is introduced to realize multi-queue task scheduling. You can set resource capacity quotas for different queues to solve quota and cost management problems in multi-tenant scenarios. Also priority scheduling. The priority of a task is usually determined by priorityClassName. Some special cases are also determined by the specifications of the resource package applied for by the task, such as the priority of the large resource package specifications, and the DRF (the task with fewer resources has a higher priority).

(4) Advanced preemption strategies, such as group preemption.

In order to solve the problem of GPU card fragmentation, binback strategy and secondary scheduler defragmentation function are needed. Community current solution: Mainly the batch scheduler Volcano. In addition, the community provides K8s-based native job queues and flexible quota manager Kueue.

(4) Topology aware scheduling

It is necessary to sense the network links between nodes and the connection mode of resources within a single node to improve task performance, speed up training, and solve the problem of inconsistent instance performance.

(5) GPU topology

It is better to schedule different instances of the same task to a combination of GPU cards with high-speed connections (nvlink, etc.). The scheduler needs to be aware of the topology of resource connections within a single node.

The following is an example of the V100 architecture:

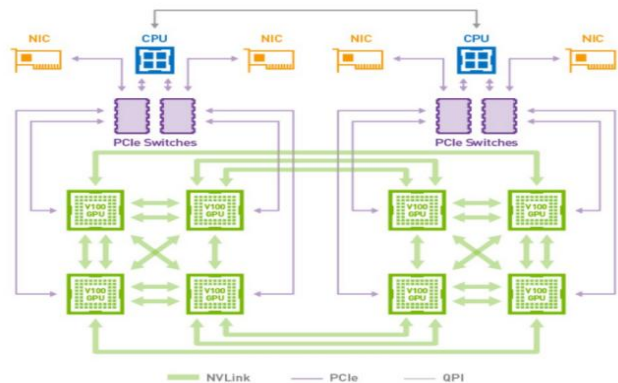

**Figure 2:** V100 automation scheduling architecture diagram

Each V100 GPU has six NVLink channels. Eight Gpus cannot be fully connected, and two Gpus can be connected to a maximum of two NVLink channels. There are two NVLink connections between GPU0 and GPU3, GPU0 and GPU4, one NVLink connection between GPU0 and GPU1, and no NVLink connection between GPU0 and 6. Therefore, GPU0 and GPU6 still need to communicate through PCIe.

As can be seen from the figure below, eight Gpus are connected through six NVswitches. Unlike the V100, there are nvlink differences between GPU cards.

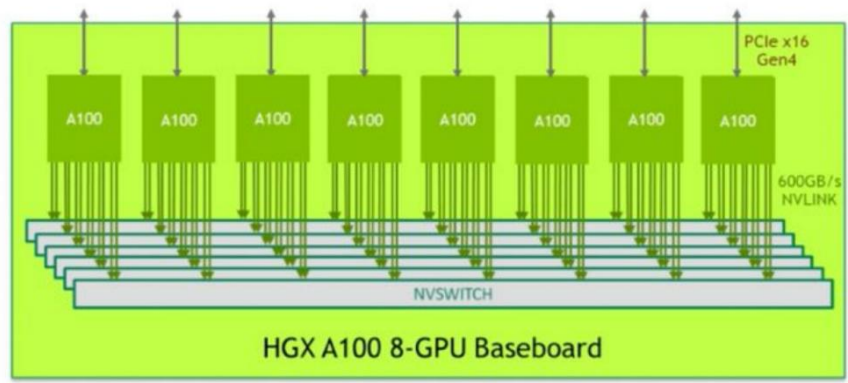

**Figure 3:** HGXA100 8-GPU Baseboard

- Network topology

Faster communication and lower latency between nodes within the same rack or tor. Mainly in the large-scale distributed training hybrid parallel scenario, the scheduler will schedule m Pods in the same unit to the Node under the same Tor as far as possible according to the convergence ratio of the switch.

**3.2 Utilization of Deep Learning**

There is a synergy between Kubernetes and deep learning in implementing automated scheduling applications. First, Kubernetes, as a container orchestration platform, uses its automatic scheduling and resource management capabilities to intelligently assign tasks according to the resource requirements of containers and the available resources of nodes to achieve efficient cluster management. Secondly, deep learning technology plays an important role in scheduling decision-making. By analyzing historical data, predicting load changes and optimizing scheduling policies, it can further improve the intelligence and adaptability of scheduling, so as to achieve more flexible and efficient task scheduling.

This combination enables the automated scheduling system to better adapt to dynamically changing workloads and make optimal scheduling decisions based on real-time performance indicators and predicted trends, thereby improving cluster utilization, reliability and performance and providing reliable infrastructure support for large-scale containerized applications.

**3.3 Integration of Reinforcement Learning**

The integration of reinforcement learning and Kubernetes automated scheduling will make the scheduling system more intelligent and adaptive. Through the reinforcement learning algorithm, the system can learn the optimal scheduling strategy according to the environmental feedback and reward signals, so as to achieve more efficient and flexible task allocation and resource management. This combination can further improve the intelligence level of the scheduling system, so that it can better adapt to the dynamic change of workload and cluster resource conditions.

Kubernetes automated scheduling has many advantages on its own, including declarative configuration, automated scheduling, fault recovery, horizontal scaling, and service discovery and load balancing capabilities. Combined with reinforcement learning techniques, these advantages can be further enhanced. For example, reinforcement learning algorithms can dynamically adjust scheduling policies based on real-time performance indicators and predicted trends to achieve more intelligent and optimized resource allocation, thereby improving cluster utilization, reliability, and performance.

**3.4 Kubernetes automated scheduling features and benefits**

(1)Declarative configuration: Kubernetes simplifies deployment and management by using declarative configuration to describe the desired state of an application rather than specifying specific action steps.

(2)Automated Scheduling: Kubernetes provides automatic scheduling capabilities that intelligently schedule based on resource requirements of containers and available resources of nodes to ensure optimal performance and resource utilization.

(3)Fault Recovery: Kubernetes has an automatic fault detection and recovery mechanism that can automatically reschedule failed container instances and ensure high availability of applications.

(4)Horizontal scaling: Kubernetes supports horizontal scaling, which automatically adjusts the number of copies of the application to meet demand as the load changes.

Service discovery and load Balancing: Kubernetes provides built-in service discovery and load balancing capabilities that make it easy for applications to implement inter-service communication and traffic distribution.

## 4. METHODOLOGY: PRACTICAL APPLICATION

**4.1 Case overview**

Let's say we have a large e-commerce platform that uses Kubernetes clusters to deploy and manage its services. During specific promotions, traffic spikes, placing extremely high demands on app availability and performance. To cope with this traffic spike, we need to ensure that Pods are scheduled efficiently and that resources are used wisely.

**4.2 Problems Encountered**

(1) Resource bottleneck: During peak traffic, some nodes respond slowly due to excessive (2) load, resulting in service interruption.
Scheduling delay: Due to the burst of high traffic, there is a significant delay in the startup and scheduling of new Pods.
(3) Unbalanced resource distribution: some nodes have overutilized resources, while others have idle resources.

**4.3 Solution**

(1) Automatic capacity expansion: Use the horizontal Pod automatic capacity expansion (HPA) and cluster automatic capacity expansion (CA) features of Kubernetes to dynamically manage resources.
(2) HPA: Automatically increase or decrease the number of Pods based on CPU and memory usage to cope with traffic changes.
(3) CA: Add more nodes when needed and reduce nodes when traffic drops to save costs.

**4.4. Optimize Pod scheduling policies**

Adjust Pod scheduling policies to ensure that Pods are evenly distributed in the cluster and avoid overloading of some nodes.
(1) Pod affinity and anti-affinity: Define appropriate affinity rules to ensure that pods of related services are distributed on different nodes to improve availability.
(2) Pod topology expansion constraints: Ensure that Pods are evenly distributed in different availability zones to avoid the failure of a single area affecting the entire service.

**4.5 Applications of the Advanced Scheduling feature**

Use Taints and Tolerations as well as custom schedulers to further optimize resource allocation.
(1) Taints and Tolerations: Set up taints for nodes that handle high traffic and only allow Pods with specific tolerations to run on these nodes.
(2) Custom scheduler: Develop a custom scheduler to optimize Pod scheduling decisions based on real-time traffic and resource usage.

**4.6 Performance Monitoring and Real-time Adjustment**

Implement a comprehensive monitoring and logging system to track cluster performance and resource usage in real time.

Monitoring tools: Use tools such as Prometheus and Grafana to monitor resource usage and service performance.

(2) Real-time adjustment: Based on monitoring data, quickly adjust scheduling policies and resource allocation to meet real-time performance requirements and resource constraints.

**4.7 Disaster Recovery and Failover**

Establish disaster recovery plans and failover mechanisms to ensure that services continue to run in the event of unforeseen problems.

(1) Multi-region deployment: The service is deployed in different geographical locations to ensure that the failure of a single region does not affect the entire platform.

(2) Fast recovery strategy: achieve fast fault detection and automated recovery process to reduce service interruption time.

**4.8 Testing and optimization**

Comprehensive testing, including stress testing and performance testing, is conducted prior to production deployment to verify the effectiveness of scheduling policies and resource allocation.

(1) Performance test: simulate high traffic conditions to test the response ability of the system and the effectiveness of resource allocation.

(2) Optimization iteration: Adjust and optimize scheduling strategy and resource allocation according to test results.

Establish feedback mechanisms to continuously collect and analyze performance data to continuously improve scheduling policies and resource management.

(3) Continuous monitoring: Implement continuous performance monitoring to ensure that any problems are detected and resolved in a timely manner.

(4) Improvement iteration: continuous scheduling strategy and resource management optimization based on collected data and feedback.

## 5. CONCLUSION

Through this research, we not only deeply analyze the application of deep learning and reinforcement learning in large-scale cloud computing systems, but also discuss the combination with Kubernetes automated scheduling. We propose a scheme that utilizes deep learning to monitor system state and reinforcement learning to adjust scheduling strategy to realize intelligent management of cluster resources and optimization of task scheduling. The experimental results show that this method has achieved remarkable results in improving system efficiency, resource utilization and performance.

Future research directions include further optimizing deep learning models to improve their ability to accurately predict system state and task characteristics, as well as improving reinforcement learning algorithms to be more flexible to different workloads and clustered environments. In addition, it should also strengthen the combination with practical application scenarios, explore more practical and feasible automatic task scheduling schemes to meet the growing demand for cloud computing, and promote the development and innovation of cloud computing technology.

In summary, the combination of deep learning and reinforcement learning technologies with Kubernetes automated scheduling brings new opportunities and challenges for large-scale cloud computing systems. The automatic task scheduling scheme proposed in this paper provides a feasible solution to improve the efficiency, resource utilization and performance of the system, and provides a useful enlightenment for the future research and practice in the field of cloud computing.